\documentclass[numbers]{article}

\usepackage[final]{neurips_2024}
\usepackage{graphicx}
\usepackage{amsmath}
\usepackage{wrapfig}
\usepackage{float}
\usepackage{listings}
\usepackage{authblk}

\usepackage{multirow}
\usepackage{subcaption}
\usepackage{geometry}
\usepackage[table,xcdraw]{xcolor}
\usepackage{caption}
\usepackage[utf8]{inputenc} 
\usepackage[T1]{fontenc}    
\usepackage{hyperref}       
\usepackage{url}            
\usepackage{booktabs}       
\usepackage{amsfonts}       
\usepackage{nicefrac}       
\usepackage{microtype}      
\usepackage{xcolor}         
\usepackage{pifont}

\usepackage{multirow}
\usepackage{siunitx}

\title{ChEmbed: Enhancing Chemical Literature Search Through Domain-Specific Text Embeddings}

\author{
\begin{minipage}[c]{\textwidth}
\centering
\textbf{
  Ali Shiraee Kasmaee\textsuperscript{1,2,*}, 
  Mohammad Khodadad\textsuperscript{1,2},
  Mahdi Astaraki\textsuperscript{1,2}, 
  Mohammad Arshi Saloot\textsuperscript{2}, 
  Nicholas Sherck\textsuperscript{3}, 
  Hamidreza Mahyar\textsuperscript{1},
  Soheila Samiee\textsuperscript{2}}\\
  \textsuperscript{1}Department of Computational Science and Engineering, McMaster University, Canada \\
  \vspace{0.2em}
  \textsuperscript{2}BASF Canada Inc., Canada\\
  \textsuperscript{3}BASF Corporation, USA\\
  \texttt{\{shiraeea, khodam3, astarakm, mahyarh\}@mcmaster.ca \linebreak
  \{mohammad.arshi-saloot, nicholas.sherck, soheila.samiee\}@basf.com} \\
  \textsuperscript{*}Corresponding Author: \texttt{shiraeea@mcmaster.ca}\\
\end{minipage}
}

\begin{document}
\raggedbottom
\maketitle

\begin{abstract}
Retrieval-Augmented Generation (RAG) systems in chemistry heavily depend on accurate and relevant retrieval of chemical literature. However, general-purpose text embedding models frequently fail to adequately represent complex chemical terminologies, resulting in suboptimal retrieval quality. Existing embedding models for chemistry are outdated, and none is tailored to chemical literature retrieval, leaving a substantial performance gap. To address this challenge, we introduce \textbf{ChEmbed}, the first purpose-built family of domain-adapted text embedding models engineered for chemical literature retrieval. These models are fine-tuned via contrastive learning on a dataset comprising chemistry-specific text from the PubChem, Semantic Scholar, and ChemRxiv corpora. To create effective training data, we employ large language models to synthetically generate queries, resulting in approximately 1.7 million high-quality query-passage pairs. Additionally, we augment the tokenizer by adding 900 chemically specialized tokens to previously unused slots, which reduces the fragmentation of chemical entities, such as IUPAC names.
ChEmbed also maintains an 8192-token context length, enabling retrieval of longer passages than many open-source embedding models allow.
Evaluated on our newly introduced \textbf{ChemRxiv Retrieval} benchmark, ChEmbed outperforms state-of-the-art general embedding models, raising MRR@10 from 0.781 to 0.882 (+10.1 pp). It also substantially outperforms domain-specific embedding models such as Chemical-BERT, improving MRR@10 from 0.096 to 0.882. A role-based retrieval analysis using PubChem descriptions and ChEBI annotations shows that the improvement extends to chemical-role queries. ChEmbed represents a practical, lightweight, and reproducible embedding solution that effectively improves chemical literature retrieval.
\end{abstract}

\section{Introduction}
Retrieval-Augmented Generation (RAG) enables large language models to leverage external sources at inference time, significantly enhancing factual accuracy when the task domain diverges from the model’s pre-training data. Chemistry epitomizes such divergence: highly specialized nomenclature, formulas, and reaction contexts rarely appear in general-purpose corpora, causing standard LLMs to hallucinate or misinterpret key concepts \cite{peng2021domain,acharya2024exploring,gu2021domain,zhang2024chemllm}. Because the quality of a RAG system is bound by the strength of its retriever, robust domain-specific text embeddings are indispensable \cite{nussbaum2024nomic}.

Despite rapid advances in universal embedding encoders, a persistent performance gap remains in chemistry. Existing scientific models still struggle with fine-grained chemical terminology, and two structural issues throttle progress: (i) \textbf{data scarcity}, since contrastive learning requires curated query-passage pairs that are costly to obtain in specialized fields (recent work explores synthetic generation as a solution \cite{adler2024nemotron,bonifacio2022inpars,jeronymo2023inpars}); and (ii) \textbf{benchmark mismatch}, as most public retrieval benchmarks rely on encyclopedic summaries that fail to capture the nuances of primary chemical literature \cite{pmlr-v262-shiraee-kasmaee24a}. Furthermore, generic sub-word tokenizers fragment chemical names and molecular descriptors, degrading semantic coherence before encoding even begins. Prior work in financial text processing shows that domain-adapted tokenizers can outperform generic subword tokenizers for representing specialized terminology and improving downstream performance \cite{bommarito2025kl3m}. Motivated by these findings, we apply a lightweight tokenizer augmentation strategy tailored to chemical nomenclature.
We address these challenges with \textbf{ChEmbed}, the first text embedding model fine-tuned specifically for chemistry literature retrieval, trained on 1.7 million synthetic query-passage pairs drawn from chemical articles and sources, and paired with a domain-adapted tokenizer that preserves key domain tokens. ChEmbed is evaluated using a new retrieval benchmark built from chemistry articles. Our study makes four contributions:
\begin{enumerate}
\item \textbf{Synthetic data for domain adaptation:} We built a pipeline for large-scale synthetic query-passage generation with LLMs for Chemical Retrieval, and demonstrate that it provides effective training data for this task.
\item \textbf{Domain-adaptive tokenizer augmentation:} A lightweight vocabulary-patching technique that injects chemistry terms into an existing WordPiece tokenizer without retraining the base tokenizer or model from scratch.
\item \textbf{Literature-driven benchmark:} We release a retrieval task benchmark constructed from chemical literature rather than encyclopedic text.
\item \textbf{ChEmbed model:} Our publicly available encoder achieves an absolute gain of 10.1 percentage points in MRR@10 over its general-purpose baseline on the chemistry-specific benchmark.
\end{enumerate}
Together, these advances narrow the domain gap for retrieval-augmented generation in chemistry, paving the way for more reliable AI systems that can accelerate chemical research and discovery.

\section{Related Work}
\paragraph*{Text Embedding Models and Domain Adaptation}
Text embedding models transform texts into fixed-length vectors capturing semantic meaning, enabling effective retrieval in tasks such as retrieval-augmented generation (RAG) \cite{wang2024multilingual}. Initially, embedding methods like Sentence-BERT (SBERT) used siamese or triplet networks to produce sentence embeddings \cite{reimers2019sentence}. Later, simpler contrastive learning approaches such as SimCSE \cite{gao2021simcse} demonstrated excellent results using minimal labels. Recent embedding models like E5 \cite{wang2022text}, Alibaba’s GTE \cite{li2023towards}, the BGE series \cite{chen2024bge}, and Qwen3 Embedding \cite{qwen3embedding} leverage large-scale pretraining and contrastive learning to create robust universal embeddings. Qwen3 Embedding notably explores decoder-only architectures, in contrast to the encoder-based approaches that have been dominant among earlier models. Notably, newer architectures such as ModernBERT \cite{warner2025smarter} and Nomic \cite{nussbaum2024nomic} have established long-context processing (8192+ tokens) as a standard, rendering the 512-token limits of earlier models insufficient for full-text scientific retrieval. Despite these advances, general embedding models often underperform in specialized domains. Researchers have responded by adapting language models specifically to the chemical domain. Some models, such as MatSciBERT \cite{gupta2022matscibert} and ChemicalBERT \cite{recobo_chemical_bert_uncased}, extend the pre-training of general-purpose encoders like BERT on chemical corpora. Other methods represent chemical structures directly, using SMILES strings \cite{wang2019smiles,zheng2024bert,chithrananda2020chemberta,ahmad2022chemberta} or molecular graphs \cite{rong2020self,liu2021pre}. However, recent evaluations on chemical RAG systems \cite{amiri2025chunk} indicate that these older domain-specific models are frequently outperformed by modern generalist retrievers. This inversion suggests that domain pre-training alone is insufficient without the massive contrastive instruction tuning employed by generalist models. While domain-specialized embedders already exist for scientific text broadly (SPECTER \cite{cohan2020specter}) and for adjacent sciences, with MedCPT \cite{jin2023medcpt} in biomedicine and PhysBERT \cite{hellert2024physbert} in physics, no released text embedding model has been trained and evaluated specifically for chemistry literature retrieval.

\paragraph*{Synthetic Data Generation}
A recent advancement in training language and retrieval models is the use of synthetic data generated by large language models (LLMs). At scale, synthetic datasets are becoming essential; for instance, NVIDIA's Nemotron relies on synthetic data generated by a 340B-parameter LLM for up to 98\% of its instruction-tuning tasks across diverse fields \cite{adler2024nemotron}. In information retrieval specifically, methods like InPars \cite{jeronymo2023inpars, bonifacio2022inpars} and Promptagator \cite{dai2022promptagator} prompt LLMs with just a few example queries to generate large amounts of synthetic queries aligned with existing passages. In the same vein, GPL \cite{wang2022gpl} adapts dense retrievers to new domains using generated queries and a cross-encoder pseudo-labeler, while Gecko \cite{lee2024gecko} distills LLM-generated query-passage pairs into a general-purpose embedder. This methodology aligns with emerging trends in semantic supervision, such as SemCSE \cite{brinner2025semcse}, which demonstrate that synthetic LLM-generated signals often yield better retrieval alignment than traditional citation-based graphs. These generated query-passage pairs enable practical training of retrieval models, significantly alleviating data scarcity. This synthetic data approach has also been successfully applied to train text-embedding models, such as \textbf{\texttt{E5-Mistral-7B-instruct}}, which demonstrated that a general-purpose model could be trained almost exclusively on synthetic query-passage pairs \cite{wang2023improving}. Likewise, specialized embedding models, such as \textbf{MedEmbed}, fine-tuned specifically for medical and clinical use cases, have effectively utilized synthetic LLM-generated data \cite{balachandran2024medembed}.

\paragraph*{Chemistry NLP Benchmarks} To evaluate text embeddings comprehensively, researchers typically rely on standard benchmarks such as BEIR and MTEB. BEIR (Benchmarking IR) is a widely-used suite containing 18 diverse retrieval datasets for zero-shot evaluation across various tasks, including question-answering and fact-checking \cite{thakur2021beir}. The Massive Text Embedding Benchmark (MTEB) further expands coverage, now including more than 100 tasks spanning over 1000 languages \cite{mteb_website} across various embedding applications like retrieval, classification, and clustering \cite{muennighoff2022mteb}. While these benchmarks are valuable for evaluating general-purpose models, they primarily feature general-domain tasks, with limited coverage of specialized scientific domains. To address this gap, domain-specific evaluation suites have been introduced. For instance, ChemTEB \cite{pmlr-v262-shiraee-kasmaee24a} provides a dedicated evaluation benchmark tailored for chemical sciences. ChemTEB encompasses 34 diverse chemical NLP tasks, including chemical text classification, clustering, text-SMILES mining, and retrieval. Although ChemTEB effectively evaluates general embedding models in chemistry contexts, it currently lacks retrieval tasks focused specifically on literature search, as its existing retrieval tasks primarily draw from encyclopedic sources. More recent chemistry benchmarks target retrieval-augmented question answering, such as ChemRAG-Bench \cite{zhong2025benchmarking}, which scores end-to-end RAG answers rather than retrieval alone. A smaller related resource, ChemRxivQuest \cite{amiri2025chemrxivquest}, provides chemistry question-answer pairs from ChemRxiv. We instead introduce a larger benchmark built to evaluate retrieval directly on held-out literature paragraphs.

\paragraph*{Tokenizer Adaptation in Domain-Specific NLP} Various WordPiece-based \cite{wu2016google} strategies have been proposed to adapt BERT tokenizers for specialized domains. \textbf{SciBERT} \cite{beltagy2019scibert} exemplifies a complete retraining approach: it replaces BERT’s entire vocabulary with a new WordPiece vocabulary derived from scientific corpora and then pre-trains a domain-specific model from scratch. While effective, this yields a high computational cost. Alternative approaches seek to bypass tokenizer modification entirely; for instance, recent multi-modal adapters like ChemLML \cite{deng2025chemical} and MoRA \cite{yin2025mora} use lightweight "linkers" to bridge text and molecular modalities. While efficient for generation, these adapter-based methods add inference complexity and do not fundamentally resolve the segmentation of chemical entities in the base encoder. In contrast, more lightweight methods inject or append domain terms into an existing model’s tokenizer without altering its architecture. For example, \textbf{CancerBERT} \cite{zhou2022cancerbert} repurposes BERT’s reserved \texttt{[UNUSED]} token slots to incorporate cancer-specific words, then continues pre-training on in-domain text, thus introducing new terminology without expanding the model size. Similarly, \textbf{AVocaDo} \cite{hong2021avocado} adapts BERT’s tokenizer by adding a small set of high-utility domain tokens to the original WordPiece vocabulary (identified from downstream data) and treating these new tokens’ embeddings as additional parameters learned during fine-tuning. Another approach, \textbf{exBERT} \cite{tai2020exbert}, extends the base tokenizer via an auxiliary embedding module: new domain-specific tokens receive their embedding vectors in a separate “extension” vocabulary, and the model learns to combine the original and extension embeddings through a trainable weighted sum, all while keeping the original BERT weights fixed. This method effectively integrates new vocabulary without modifying the core model, albeit with added implementation complexity. Overall, these approaches illustrate a spectrum of trade-offs between complete re-training for a custom vocabulary and more practical vocabulary augmentation techniques for domain adaptation. Concurrent with our work, Kalamkar et al. \cite{kalamkar2025tokenization} extend a language model's vocabulary with chemistry tokens (including SMILES sequences) and continue pretraining to improve chemical representation; by contrast, we inject chemistry tokens into the reserved \texttt{[UNUSED]} slots of a retrieval encoder without any additional pretraining.

\section{Dataset Construction}
Practical training of bi-encoder-based text embedding models typically relies on structured data in the form of (query, passage) pairs or triplets (query, positive passage, negatives), particularly when employing contrastive learning objectives \cite{li2023towards,xiao2024c,wang2022text,nussbaum2024nomic}. However, such structured datasets are often not readily available in specialized domains, such as chemistry. To enable the training of a chemistry-adapted embedding model, we begin by collecting paragraphs from chemistry-related scientific literature. Once these domain-specific passages are gathered, we use a Large Language Model (LLM), carefully prompted, to generate relevant queries corresponding to each paragraph. This process enables us to construct the paired data necessary for practical contrastive training in the chemistry domain. The datasets built and used in this study are summarized in Table~\ref{tab:chemistry-datasets}.

\begin{table}[t]
\caption{Summary of chemistry domain datasets used for training and evaluation}
\label{tab:chemistry-datasets}
\centering
\resizebox{\textwidth}{!}{%
\begin{tabular}{c l l c c c}
\toprule
Idx & Dataset              & Description                                                     & \# Samples & Usage & LLM used for query synthesis \\
\midrule
1   & PubChem compounds    & Title, IUPAC, SMILES, synonyms & 2,087,164 & Tokenizer & N/A \\
2   & PubChem descriptions &  Compound descriptions    &   393,321  & Training & \texttt{gpt-4.1-nano} \\
3   & S2ORC chemistry      & Filtered chemistry paragraphs  & 1,187,726  & Training & \texttt{gpt-4.1-nano} \\
4   & ChemRxiv paragraphs       & Extracted ChemRxiv paragraphs (CC-BY)               &   139,057 & Training & \texttt{o3-mini} \\
5   & ChemRxiv paragraphs        & Extracted ChemRxiv paragraphs (CC-BY-NC) &    69,457  & Evaluation & \texttt{Claude 3.7 Sonnet} \\
6   & ChemRxiv metadata    & Title and abstract ChemRxiv preprints           &    30,378 & Training & N/A \\
\bottomrule
\end{tabular}}
\end{table}

\subsection{Data Sources and Preprocessing}
We used the following sources to gather chemistry-related paragraphs:
\begin{enumerate}
\item \textbf{PubChem} is a comprehensive database of chemical entities that contains information on over 100 million unique compounds, including names, SMILES, IUPAC names, synonyms, molecular formulas, and descriptions; it offers programmatic access (PUG-View) \cite{kim2019pubchem} through which we collected data, many of which lacked descriptions, resulting in around 393 thousand usable descriptions after preprocessing.
\item \textbf{S2ORC} is an extensive collection of over 81 million academic papers, 8.1 million of which include structured full text \cite{lo2019s2orc, peS2o}. This dataset spans multiple disciplines and provides rich metadata, citation information, and full-text annotations. We specifically used the portion of the corpus licensed under the public domain and Creative Commons BY, comprising 6.2 million papers. From this subset, we extracted approximately 118,000 documents tagged with the chemistry subject, split them into individual paragraphs, and, after filtering out conclusion sections, table/figure captions, and short or uninformative segments, retained around 1.18 million high-quality paragraphs.
\item \textbf{ChemRxiv} is a free, open-access preprint server for chemistry and related fields, offering early distribution of research findings. It provides a public API, which we utilized to extract metadata and PDFs for approximately 30 thousand chemistry manuscripts. We processed these documents using GROBID \cite{GROBID}, a tool for extracting structured information from scientific publications, and segmented the texts into individual paragraphs. To ensure data quality, following a methodology similar to that used in the S2ORC dataset \cite{lo2019s2orc, peS2o}, we filtered out paragraphs with an average unigram log probability less than -20 or containing fewer than 50 words. This preprocessing resulted in a corpus of approximately 209 thousand high-quality paragraphs.
\end{enumerate}
\subsection{Synthetic Query Generation via LLMs}
\label{subsec:query-generation}
To create high-quality paired data for contrastive training, we leveraged Large Language Models (LLMs) to generate synthetic queries directly from chemistry paragraphs. Our goal was to closely replicate realistic information retrieval scenarios, such as a user typing a specific, chemistry-focused query into a search system to retrieve a relevant passage that answers the question. To achieve this, we carefully designed an LLM prompt instructing the model to produce exactly one clear, meaningful chemistry question that the given paragraph can answer. The prompt explicitly disallowed superficial or yes/no questions, as well as references to the text itself (e.g., "according to this paragraph"). We used two LLMs from the OpenAI platform (\texttt{o3-mini} and \texttt{gpt-4.1-nano}) chosen to balance scale, cost, and data complexity during synthetic query generation. For constructing the evaluation retrieval dataset, we used a separate model, \texttt{Claude 3.7 Sonnet} with extended thinking, applied to held-out ChemRxiv paragraphs not seen during training.

After applying this procedure, our dataset consisted only of the query-passage pairs that strictly met our generation criteria. The LLMs refused to generate queries for approximately 29 thousand paragraphs, effectively filtering them out during the process. Upon manual inspection of a sample of these refused cases, we confirmed that the LLM consistently excluded paragraphs that were irrelevant, too brief, or lacked meaningful scientific content, such as funding acknowledgments, overly general conclusions, or short and information-poor text segments. As a post-hoc quality check, one reviewer assessed 50 query-passage pairs used in model training and 50 retrieval cases used in model evaluation, as described in Section~\ref{subsec:expert-review}.

\begin{figure}[!b]
        \centering
        \includegraphics[width=\linewidth]{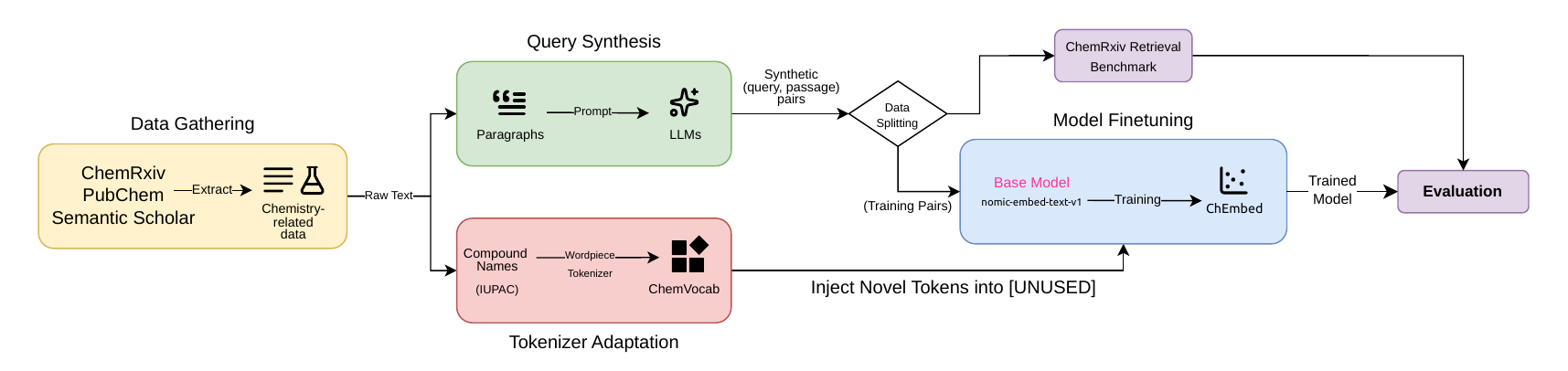}
        \caption{Overview of the ChEmbed pipeline. Chemistry-related paragraphs are extracted from multiple sources and paired with synthetic queries generated by prompting large language models (LLMs). A held-out subset of ~69,000 paragraphs is reserved for evaluation, with distinct LLMs used for generating training and evaluation queries. In parallel, a ChemVocab tokenizer is trained from more than two million IUPAC names, and 900 unique tokens are injected into the \texttt{bert-base-uncased} unused tokens. The base encoder (nomic-embed-text-v1) is then fine-tuned on the generated query-passage pairs to produce the final ChEmbed family of models.}
        \label{fig:pipeline}
\end{figure}

\subsection{Expert Review of Synthetic Queries and Retrieval Results}
\label{subsec:expert-review}
\begin{figure}[!t]
    \centering
    \includegraphics[width=\linewidth]{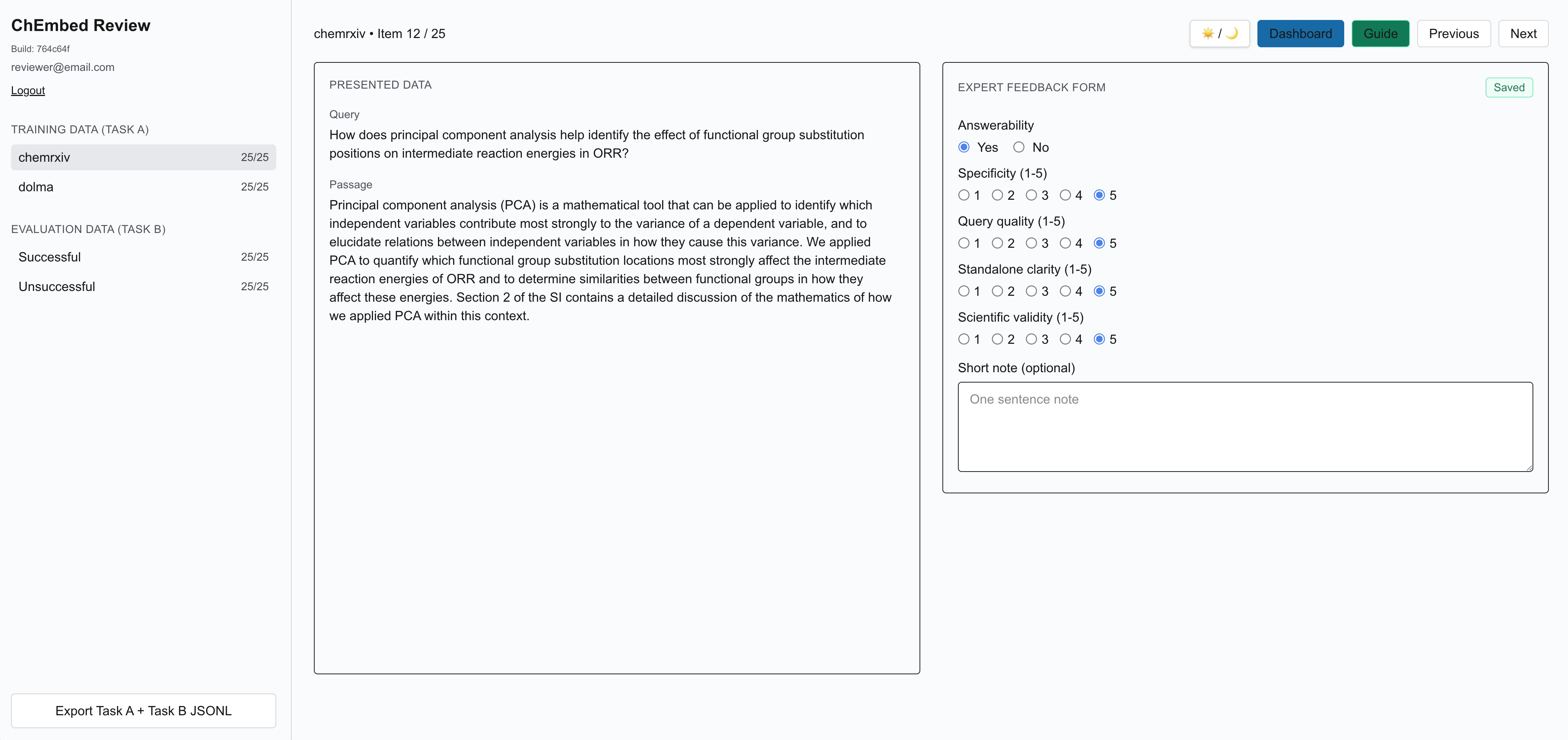}
    \caption{Expert-review interface for Task A. The reviewer inspected a generated query alongside its real source chemistry passage and recorded answerability, specificity, query quality, standalone clarity, scientific validity, and an optional qualitative note.}
    \label{fig:expert-review}
\end{figure}

To assess whether the generated queries represented meaningful chemistry information needs, we conducted a structured review with one domain expert annotator (with a post-graduate degree in Chemistry). The review comprised two tasks and used the interface shown in Fig.~\ref{fig:expert-review}. Task A covered 50 training query-passage pairs, with 25 from ChemRxiv and 25 from the Dolma chemistry collection. For each pair, the reviewer judged whether the source passage answered the query and assigned ratings from 1 (low) to 5 (high) for specificity, query quality, standalone clarity, and scientific validity. Specificity reflected how closely the query was tied to the passage; query quality, its naturalness and usefulness as a search query; standalone clarity, whether it could be understood without additional context; and scientific validity, its faithfulness to the chemistry in the passage. An optional free-text note was also available.

In the evaluation, only the passage used to generate each query was considered relevant, and MRR was calculated from the rank of that passage. To evaluate whether other correct documents might exist for the queries, a second task was defined.
In this task, we selected 50 cases from the ChemRxiv evaluation set, evenly split between 25 cases in which the designated gold passage appeared in the top 10 results and 25 cases in which it did not. The reviewer saw the query, its designated gold passage, and the top 10 passages retrieved by \texttt{ChEmbed\textsubscript{progressive}}. In addition to evaluating the queries and answers using the Task A criteria, the human expert was asked to classify each non-gold passage as either \textit{not relevant}, \textit{somewhat relevant}, or \textit{relevant}. If a strongly relevant non-gold passage was identified, the reviewer could mark it as a near miss.

For Task A, 48 of 50 queries (96\%) were judged answerable. The mean ratings were 4.76 for specificity, 4.56 for query quality, 4.92 for standalone clarity, and 4.72 for scientific validity. For Task B, 45 of 50 queries (90\%) were judged answerable, with corresponding mean ratings of 4.98, 4.84, 5.00, and 5.00. At least one near miss (a strongly relevant non-gold passage) was identified in 32 of the 50 retrieval cases. No item was removed from the evaluation set as a result of this review, and its findings did not feed back into any evaluation.

We used these annotations as a post-hoc human assessment of synthetic-query quality and as a diagnostic complement to the automatic rank-based metrics. The review outcomes were not used to filter or regenerate training pairs, update model parameters, or alter the benchmark relevance labels.

\section{Model Architecture and Domain Adaptation}
A core challenge in building domain-specialized embedding models is choosing a base architecture that is both performant and open for further adaptation. For this work, we selected the nomic embedding family \cite{nussbaum2024nomic}, which represents one of the best and most lightweight open-source text embedding models available, supporting long context up to 8192 tokens. Unlike many proprietary or partially open alternatives, the nomic models provide full access to intermediate weights and pretraining data, enabling reproducibility and deeper scientific analysis. Architecturally, nomic builds on a BERT but introduces key architectural improvements, including rotary positional embeddings, FlashAttention for efficient long-context processing, and SwiGLU activations \cite{shazeer2020glu}, that collectively improve both accuracy and scalability. It is also the best-performing open-source embedding model on the domain-specific chemical benchmark \cite{pmlr-v262-shiraee-kasmaee24a}, and its training on large general corpora already gives it a broad knowledge of chemical language. For these reasons, nomic models serve as an ideal starting point for domain adaptation in chemistry.

\subsection{Domain Adaptation Strategies}
To adapt the nomic base model for chemistry retrieval, we explored several fine-tuning strategies with both supervised and unsupervised objectives. Two key variants were considered, both originally trained by the nomic team: \texttt{nomic-embed-text-v1-unsupervised}, initially trained on 235 million unsupervised pairs and a max learning rate of \num{2e-4}, and \texttt{nomic-embed-text-v1}, which was further fine-tuned on 1.6 million hard-negative mined supervised triplets with a lower learning rate of \num{2e-5}.

For model training, we considered two data formats: (1) direct use of query-passage pairs, and (2) constructing triplets (query, document, negatives). Both variants optimize the same contrastive InfoNCE objective \cite{oord2018representation}:
\[
\mathcal{L}
= -\frac{1}{N}\sum_{i=1}^{N}
\log
\frac{e^{\,s(q_i,d_i^+)/\tau\,}}
{\,e^{\,s(q_i,d_i^+)/\tau\,}
  +\displaystyle\sum_{d^-\in\mathcal{N}(q_i)}e^{\,s(q_i,d^-)/\tau\,}
}\!
\]
where \(s(q,d)\) is the cosine similarity scaled by temperature \(\tau\), \(d_i^+\) is the true positive document for query \(q_i\), and \(\mathcal{N}(q_i)\) denotes the set of negatives for \(q_i\). In the “pairs” configuration, \(\mathcal{N}(q_i)\) contains all other passages in the batch (in-batch negatives), while in the “triplets” configuration, it contains exactly the \(H\) pre-mined negatives we attach to each query. Following the nomic setup, we used 7 negatives per query, as increasing beyond this showed little additional improvement. We experimented with three approaches (7 hard, 7 random, and 3 hard + 4 random negatives). Our experiments showed that fine-tuning with pure in-batch negatives outperformed the triplet-style objective on our synthetic chemistry data, likely because our “hard” negatives were model-generated rather than human-labeled. Most hyperparameters were kept the same as those reported for nomic, with changes including a linear warmup over 5\% of the total steps, and training with a maximum context length of 2048 tokens (enabling the model to process extended sequences of up to 8192 tokens during inference by dynamically scaling the positional embeddings \cite{emozilla2023ntk,peng2023yarn}). Models were trained on 4×NVIDIA A100 40GB GPUs using GradCache \cite{gao2021scaling} and mixed precision \cite{micikevicius2017mixed}, which enabled the use of a large total batch size of \textbf{$16,384$} for more effective contrastive learning.

\subsection{Tokenizer Adaptation}
A persistent limitation in adapting language models to chemistry is the suboptimal tokenization of complex nomenclature (e.g., IUPAC names, SMILES). Like many other open-source text embedding models built on BERT \cite{devlin2019bert}, the first generation of nomic embedding models utilizes the standard \textbf{\texttt{bert-base-uncased}} tokenizer, which has a vocabulary size of 30,522. Notably, this tokenizer includes exactly 994 [UNUSED] tokens, reserved placeholders not assigned to any word or subword in the original vocabulary. When extending a model's tokenizer with fewer than 994 new tokens, a minimal intervention strategy is possible: simply repurpose the unused slots to encode new, domain-specific terms. This "plug-and-play" approach preserves the original tokenizer’s structure and compatibility while directly enhancing its ability to represent specialized chemical language. To construct a chemistry-adapted tokenizer, we trained a WordPiece tokenizer \cite{wu2016google} on more than \textbf{2 million} IUPAC names from PubChem compounds. Tokens already present in \texttt{bert-base-uncased} were removed, and 900 chemistry tokens were then injected into the \textbf{\texttt{[UNUSED]}} slots to create our adapted tokenizer \textbf{ChemVocab}. Of these 900 tokens, a curated subset consists of IUPAC nomenclature fragments identified with the rule-based IUPAC tokenizer of Krasnov \textit{et al.} \cite{krasnov2021transformer}, while the remaining trained tokens were selected with preference for entries already present in SciBERT but absent from \texttt{bert-base-uncased} \cite{beltagy2019scibert}. Table~\ref{tab:chemvocab-examples} provides a qualitative comparison between \textbf{ChemVocab} and the standard \texttt{bert-base-uncased} tokenizer, demonstrating how our domain-specific augmentation preserves the semantic integrity of chemical nomenclature by preventing excessive subword fragmentation. For these newly injected tokens, we initialized their embeddings by sampling from a normal distribution with a mean of 0 and a standard deviation of 0.2. This procedure enabled the model to encode rich chemical terminology with minimal architectural changes. A summary of the complete data and model pipeline is illustrated in Fig.~\ref{fig:pipeline}.
\begin{table}[t]
\centering
\small
\setlength{\tabcolsep}{6pt}
\renewcommand{\arraystretch}{1.15}
\caption{Qualitative comparison of chemical nomenclature segmentation: Our augmented \textbf{ChemVocab} maintains semantic sub-units (e.g., functional groups) compared to the over-segmentation observed in the general-purpose \texttt{bert-base-uncased} tokenizer.}
\label{tab:chemvocab-examples}
\begin{tabular}{@{}l l l@{}}
\toprule
\textbf{Term} & \textbf{ChemVocab (ours)} & \textbf{\texttt{bert-base-uncased}} \\
\midrule
\texttt{chlorofluorocarbon}     & \texttt{chloro/fluoro/carbon}            & \texttt{ch/lor/of/lu/oro/carbon} \\
\texttt{acetaminophen}          & \texttt{acet/amino/phen}                 & \texttt{ace/tam/ino/ph/en} \\
\texttt{secbutylamine}          & \texttt{sec/butyl/amine}                  & \texttt{sec/bu/ty/lam/ine} \\
\texttt{nitrophenylphosphate}   & \texttt{nitrophenyl/phosphate}           & \texttt{ni/tro/ph/en/yl/ph/os/phate} \\
\bottomrule
\end{tabular}
\end{table}

\section{Experiments \& Results}
\label{sec:exp-results}
\paragraph*{Experimental setup} We fine-tune \texttt{nomic-embed-text-v1} for the chemical retrieval task. The fine-tuning is performed with and without adding new tokens to the tokenizer. For the optimized addition of new tokens, we tried three approaches, which are explained in more detail in Section~\ref {subsec:tokenizer} and Table~\ref{tab:tok-ablation}.

Effectiveness is assessed through retrieval tasks using three benchmark suites: ChemRxiv Retrieval, MTEB (English v2), and ChemTEB. {ChemRxiv Retrieval} is our newly developed benchmark and, to our knowledge, the largest retrieval benchmark built directly from chemistry literature paragraphs, comprising a corpus of 69,457 paragraphs from chemical literature published on ChemRxiv and 5,000 synthetic queries. Queries are generated using a Large language model (Anthropic's Claude 3.7 Sonnet) different from the one used for training, thereby mitigating potential generation bias. This task has been integrated into the latest release of ChemTEB. {MTEB (English v2)} \cite{muennighoff2022mteb} is a widely recognized retrieval benchmark featuring 41 English-language tasks across seven categories, including classification, clustering, and retrieval. However, it is a general domain benchmark and does not focus on chemistry use cases. Given that our model is English and the primary goal of this work is to adapt models for retrieval tasks in the chemical domain, we exclusively evaluate the model using its retrieval datasets. \textbf{ChemTEB} \cite{pmlr-v262-shiraee-kasmaee24a} is a chemistry-focused adaptation of MTEB containing two retrieval tasks mainly sourced from encyclopedic data, aiming to evaluate the generalizability of embedding models to chemical domains.

\subsection{Domain-Specific Retrieval Performance}
\begin{table}[H]
    \centering\small
    \caption{Impact of different tokenizer-adaptation variants on ChemRxiv retrieval performance}
    \label{tab:tok-ablation}
    \begin{tabular}{lcc}
    \toprule
    \textbf{Variant} & $MRR@10$ & $\Delta$ vs.\ baseline (pp) \\
    \midrule
    \texttt{nomic-embed-text-v1} (baseline)       & 0.781 & --    \\
    \texttt{ChEmbed\textsubscript{vanilla}}       & 0.879 & +9.8  \\
    \texttt{ChEmbed\textsubscript{full}}          & 0.869 & +8.8  \\
    \texttt{ChEmbed\textsubscript{plug}}          & 0.878 & +9.7  \\
    \textbf{\texttt{ChEmbed\textsubscript{progressive}}} & \textbf{0.882} & +10.1 \\
    \bottomrule
    \end{tabular}
\end{table}

\paragraph*{Quantitative performance comparison}
Table~\ref{tab:chemrxiv-results} presents the performance of different model variants evaluated on the ChemRxiv Retrieval benchmark. Our ChEmbed variants consistently outperform not only the initial \texttt{nomic-embed-text-v1} baseline, but also all other notable open-source and proprietary embedding models evaluated to date.
The vanilla model \textbf{ChEmbed\textsubscript{vanilla}}, which uses the original BERT tokenizer, achieved an \textit{MRR@10} of \textbf{0.879}, representing a +9.8 percentage point gain over the \texttt{nomic-embed-text-v1} baseline (0.781). Our best-performing variant, which utilizes a progressive tokenizer adaptation schedule (\textbf{ChEmbed\textsubscript{progressive}}), achieves an \textit{MRR@10} of \textbf{0.882}, representing a total improvement of +10.1 percentage points over the base model. All ChEmbed variants also outperform the strongest alternative open-source model, \texttt{Qwen3-Embedding-8B}, which reaches only 0.846 \textit{MRR@10} despite being over 55 times larger in parameter count.

\begin{figure}[H]
        \centering
        \includegraphics[width=\linewidth]{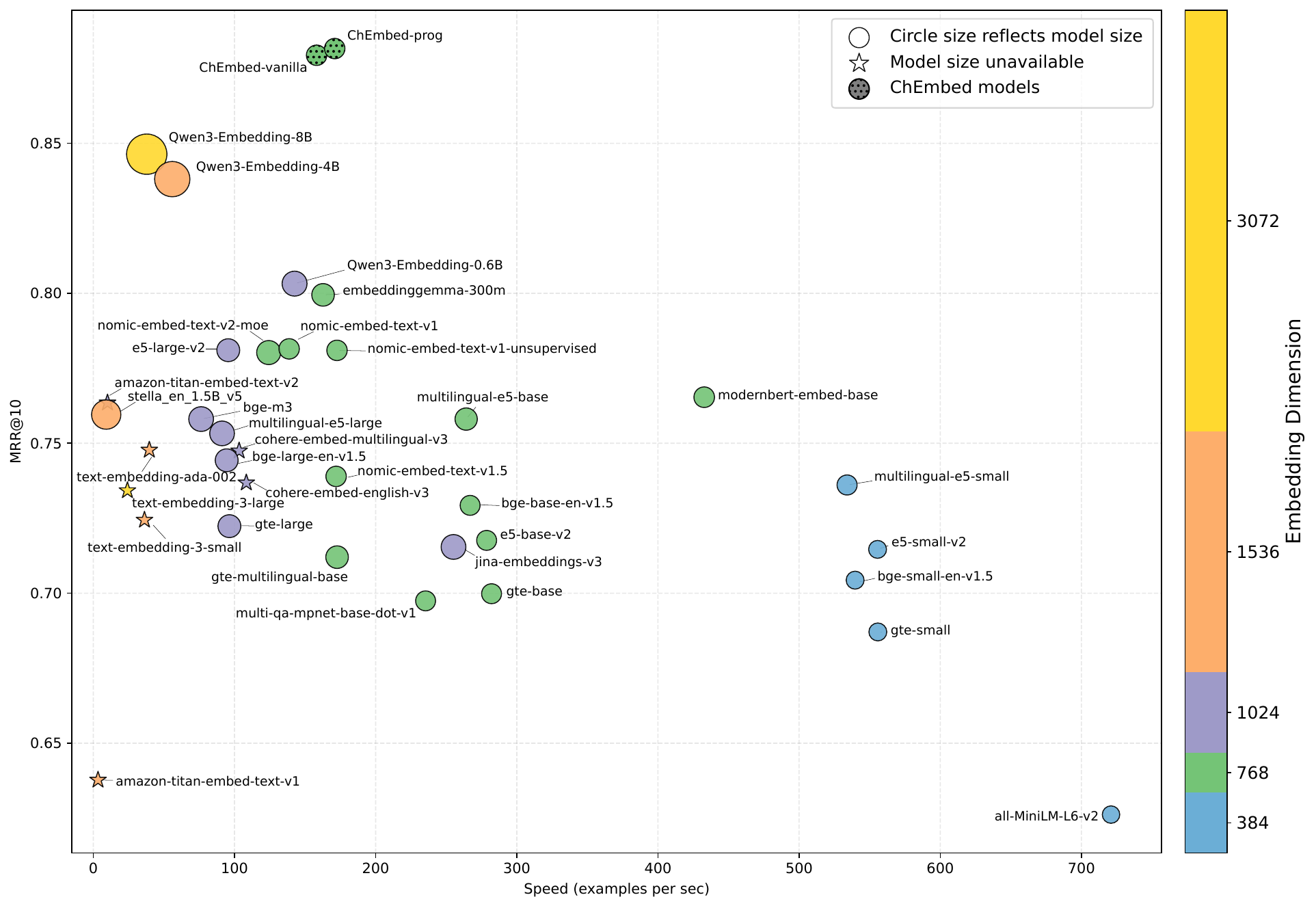}
        \caption{Model efficiency on the ChemRxiv retrieval task. The x-axis represents retrieval speed, the y-axis shows ranking effectiveness measured by Mean Reciprocal Rank ($MRR@10$), and circle sizes represent the number of model parameters.}
        \label{fig:bubble}
\end{figure}

\paragraph*{Speed-performance trade-off analysis}
Fig.~\ref{fig:bubble} visualizes the efficiency-performance balance of various embedding models on the ChemRxiv retrieval. The models are plotted according to retrieval speed (examples per second) on the horizontal axis and \textit{MRR@10} on the vertical axis. The marker size indicates the model's parameter count, while color reflects the maximum embedding dimension provided by the model. On an NVIDIA A10 24GB GPU, \textbf{ChEmbed\textsubscript{vanilla}} and \textbf{ChEmbed\textsubscript{progressive}} processed approximately 158 and 171 examples per second, respectively (a mean of 165), whereas \texttt{Qwen3-Embedding-4B} and \texttt{Qwen3-Embedding-8B} processed approximately 56 and 38 examples per second, respectively. The two Qwen3 models contain approximately 29 and 55 times as many parameters as ChEmbed, respectively, yet still fall short in retrieval accuracy, with their best variant (\texttt{Qwen3-Embedding-8B}) achieving only 0.846 \textit{MRR@10} compared to ChEmbed's 0.882. This demonstrates that ChEmbed achieves state-of-the-art chemical literature retrieval performance while being significantly faster and more lightweight than its closest open-source alternatives.

\begin{table}[!p]
\centering \small
\caption{Performance of embedding models on the \textbf{ChemRxiv Retrieval} task. ChemRxiv Retrieval has exactly one relevant document per query, so MAP@10 and MRR@10 coincide; we report MRR@10 only. “N/A” means the provider has not released parameter counts. Best scores per metric are shown in bold.}
\label{tab:chemrxiv-results}
\begin{tabular}{lcccc}
\toprule
Model Name & Emb. size & \#Params (M) & MRR@10 & NDCG@10 \\
\midrule
    \multicolumn{5}{l}{\textbf{Open-Source Models}}\\
    \texttt{chemical-bert-uncased} & 768 & 110.0 & 0.096 & 0.110 \\
    \texttt{matscibert} & 768 & 110.0 & 0.117 & 0.137 \\
    \texttt{nomic-bert-2048} & 768 & 137.3 & 0.019 & 0.025 \\
    \texttt{ModernBERT-base} & 768 & 149.7 & 0.047 & 0.056 \\
    \texttt{ModernBERT-large} & 1024 & 395.9 & 0.048 & 0.058 \\
    \texttt{scibert\_scivocab\_uncased} & 768 & N/A & 0.100 & 0.119 \\
    \texttt{bert-base-uncased} & 768 & 110.1 & 0.099 & 0.117 \\
    \texttt{all-MiniLM-L12-v2} & 384 & 33.4 & 0.556 & 0.603 \\
    \texttt{all-MiniLM-L6-v2} & 384 & 22.7 & 0.626 & 0.674 \\
    \texttt{all-mpnet-base-v2} & 768 & 109.0 & 0.618 & 0.670 \\
    \texttt{multi-qa-mpnet-base-dot-v1} & 768 & 109.5 & 0.697 & 0.741 \\
    \texttt{e5-small} & 384 & 33.0 & 0.682 & 0.726 \\
    \texttt{e5-base} & 768 & 109.0 & 0.728 & 0.770 \\
    \texttt{e5-large} & 1024 & 335.0 & 0.765 & 0.806 \\
    \texttt{e5-small-v2} & 384 & 33.0 & 0.715 & 0.756 \\
    \texttt{e5-base-v2} & 768 & 109.0 & 0.718 & 0.761 \\
    \texttt{e5-large-v2} & 1024 & 335.0 & 0.781 & 0.821 \\
    \texttt{multilingual-e5-small} & 384 & 118.0 & 0.736 & 0.778 \\
    \texttt{multilingual-e5-base} & 768 & 278.0 & 0.758 & 0.798 \\
    \texttt{multilingual-e5-large} & 1024 & 560.0 & 0.753 & 0.794 \\
    \texttt{gte-small} & 384 & 33.4 & 0.687 & 0.735 \\
    \texttt{gte-base} & 768 & 109.5 & 0.700 & 0.748 \\
    \texttt{gte-large} & 1024 & 335.1 & 0.722 & 0.768 \\
    \texttt{gte-multilingual-base} & 768 & 305.0 & 0.712 & 0.761 \\
    \texttt{bge-small-en} & 384 & 33.4 & 0.705 & 0.751 \\
    \texttt{bge-base-en} & 768 & 109.0 & 0.721 & 0.766 \\
    \texttt{bge-large-en} & 1024 & 335.0 & 0.741 & 0.785 \\
    \texttt{bge-small-en-v1.5} & 384 & 33.4 & 0.704 & 0.750 \\
    \texttt{bge-base-en-v1.5} & 768 & 109.0 & 0.729 & 0.773 \\
    \texttt{bge-large-en-v1.5} & 1024 & 335.0 & 0.744 & 0.787 \\
    \texttt{bge-m3} & 1024 & 568.0 & 0.758 & 0.798 \\
    \texttt{nomic-embed-text-v1-unsupervised} & 768 & N/A & 0.781 & 0.821 \\
    \texttt{nomic-embed-text-v1} & 768 & N/A & 0.781 & 0.820 \\
    \texttt{nomic-embed-text-v1.5} & 768 & 137.0 & 0.739 & 0.783 \\
    \texttt{nomic-embed-text-v2-moe} & 768 & 475.3 & 0.780 & 0.819 \\
    \texttt{modernbert-embed-base} & 768 & 149.0 & 0.765 & 0.807 \\
    \texttt{stella\_en\_1.5B\_v5} & 1536 & 1540.0 & 0.760 & 0.802 \\
    \texttt{jina-embeddings-v3} & 1024 & 572.0 & 0.715 & 0.759 \\
    \texttt{embeddinggemma-300m} & 768 & 307.6 & 0.799 & 0.836 \\
    \texttt{Qwen3-Embedding-0.6B}$^{a}$ & 1024 & 595.8 & 0.803 & 0.840 \\
    \texttt{Qwen3-Embedding-4B}$^{a}$ & 2560 & 4021.8 & 0.838 & 0.871 \\
    \texttt{Qwen3-Embedding-8B}$^{a}$ & 4096 & 7567.3 & 0.846 & 0.878 \\
    \texttt{ChEmbed\textsubscript{vanilla}} & 768 & 136.7 & 0.879 & 0.903 \\
    \texttt{ChEmbed\textsubscript{progressive}} & 768 & 136.7 & \textbf{0.882} & \textbf{0.906} \\
    \midrule
    \multicolumn{5}{l}{\textbf{Proprietary Models}}\\
    \texttt{text-embedding-ada-002} & 1536 & N/A & 0.748 & 0.790 \\
    \texttt{text-embedding-3-small} & 1536 & N/A & 0.724 & 0.771 \\
    \texttt{text-embedding-3-large} & 3072 & N/A & 0.734 & 0.780 \\
    \texttt{amazon-titan-embed-text-v1} & 1536 & N/A & 0.638 & 0.689 \\
    \texttt{amazon-titan-embed-text-v2} & 1024 & N/A & 0.763 & 0.805 \\
    \texttt{cohere-embed-english-v3} & 1024 & N/A & 0.737 & 0.781 \\
    \texttt{cohere-embed-multilingual-v3} & 1024 & N/A & 0.747 & 0.789 \\
    \bottomrule
\multicolumn{5}{l}{\footnotesize $^{a}$\,Loaded the model with bfloat16 to fit into GPU VRAM.}\\\end{tabular}
\end{table}

\subsection{Tokenizer adaptation analysis}
\label{subsec:tokenizer}

For the chemical retrieval task, we compared four tokenizer configurations while keeping the rest of the fine-tuning recipe unchanged:
\begin{enumerate}
\item \textbf{ChEmbed\textsubscript{vanilla}}: The vanilla variant with \texttt{bert-base-uncased} tokenizer and no vocabulary augmentation.
\item \textbf{ChEmbed\textsubscript{full}}: ChemVocab tokenizer, \emph{all} embedding parameters are trainable.
\item \textbf{ChEmbed\textsubscript{plug}}: ChemVocab tokenizer, but only the new token embeddings are updated while original BERT token embeddings stay frozen; the rest of the network is trainable.
\item \textbf{ChEmbed\textsubscript{progressive}}: progressive schedule, first train some epochs only the new token embeddings (all other weights frozen), then unfreeze and train the whole network for additional epochs.
\end{enumerate}

To assess the impact of tokenizer modifications on retrieval effectiveness, Table~\ref{tab:tok-ablation} compares the four ChEmbed configurations with the base model. All fine-tuned variants improve substantially on the target benchmark, but the differences among tokenizer configurations are small and non-monotonic. The progressive adaptation strategy achieves the highest score, although its improvement over vanilla is modest; Section~\ref{subsec:chebi-role-retrieval} examines the same two checkpoints on a role-based retrieval task, where the difference between them is more pronounced. These results suggest that the training schedule can help integrate the adapted vocabulary, but they do not establish that vocabulary augmentation consistently improves retrieval or preserves general-domain performance.

\begin{table}[H]
\centering \small
\caption{Performance comparison across shared non‑retrieval task categories in MTEB and ChemTEB, covering open‑domain and chemistry‑specific data, respectively. Evaluation metrics are: accuracy for Classification, V‑measure for Clustering, and average precision (AP) for Pair Classification. Mean (Task) reports the average score across all individual tasks, while Mean (Task Type) first averages results within each task category and then computes the mean across the category‑level averages.
}
\label{tab:nonretrieval-wide}
\setlength{\tabcolsep}{5pt}
\resizebox{\textwidth}{!}{
\begin{tabular}{l
              ccc cc
              ccc cc}
\toprule
& \multicolumn{5}{c}{\textbf{ChemTEB}} & \multicolumn{5}{c}{\textbf{MTEB}}\\
\cmidrule(lr){2-6}\cmidrule(lr){7-11}
Model
& Cls & Clust & Pair & Mean (T) & Mean (T-type)
& Cls & Clust & Pair & Mean (T) & Mean (T-type)\\
\midrule
nomic-embed-text-v1-unsupervised & 0.825 & 0.560 & \textbf{0.632} & 0.763 & 0.672
               & 0.688 & 0.442 & 0.832 & 0.607 & 0.654 \\
nomic-embed-text-v1 & \textbf{0.838} & \textbf{0.607} & 0.591 & \textbf{0.768} & \textbf{0.679}
               & 0.684 & \textbf{0.458} & \textbf{0.851} & \textbf{0.616} & \textbf{0.665} \\
\texttt{ChEmbed\textsubscript{vanilla}} & 0.807 & 0.508 & 0.585 & 0.736 & 0.633
               & 0.698 & 0.434 & 0.846 & 0.610 & 0.659 \\
\texttt{ChEmbed\textsubscript{full}} & 0.809 & 0.509 & 0.550 & 0.730 & 0.623
               & 0.698 & 0.433 & 0.844 & 0.609 & 0.658 \\
\texttt{ChEmbed\textsubscript{plug}} & 0.806 & 0.477 & 0.550 & 0.725 & 0.611
               & 0.696 & 0.430 & 0.843 & 0.607 & 0.656 \\
\texttt{ChEmbed\textsubscript{progressive}} & 0.817 & 0.497 & 0.545 & 0.733 & 0.619
               & \textbf{0.699} & 0.437 & 0.849 & 0.613 & 0.662 \\
\bottomrule
\end{tabular}
}
\end{table}

\subsection{Impact of Task and Domain Adaptation}
\label{subsec:adaptation-impact}
The goal of this work is chemical literature search, and the training data reflect that focus: fine-tuning used 1.7 million retrieval-style query--passage pairs over chemistry text, more than 98\% of which pair an LLM-generated query with a real chemistry passage. Because specialization through fine-tuning can alter performance across tasks and data distributions \citep{mosbach2023analyzing}, we evaluated all model variants on two public suites beyond our target benchmark: ChemTEB \cite{pmlr-v262-shiraee-kasmaee24a}, which covers chemistry-related embedding tasks, and MTEB English v2 \citep{muennighoff2022mteb}, a general-purpose benchmark spanning multiple tasks and domains.

The non-retrieval results are consistent with this specialization. As Table~\ref{tab:nonretrieval-wide} shows, the base model generally retains an advantage on classification, clustering, and pair-classification tasks in both the chemistry-specific and general-purpose suites. The ChEmbed variants therefore trade some performance on objectives absent from their training data for improved performance on the target literature-retrieval task. These results delimit the intended use of the models: ChEmbed was optimized for retrieval rather than as a universal replacement for the base model across embedding tasks.

The retrieval comparison in Table~\ref{tab:retrieval-wide} warrants closer attention. Fine-tuning raises ChemRxiv Retrieval \textit{MRR@10} from 0.781 to 0.879 for ChEmbed\textsubscript{vanilla}. On the two encyclopedic ChemTEB tasks, which are chemistry-related subsets of Natural Questions and HotpotQA \cite{kwiatkowski2019natural,yang2018hotpotqa}, the mean changes from 0.728 to 0.672. The ChemTEB(latest) score averages all three tasks, so it combines the ChemRxiv gain with the encyclopedic losses. The two encyclopedic tasks are also small, with 27 and 18 test queries. A single query moving from first to second place changes MRR@10 by about 0.02 on ChemNQ and 0.03 on ChemHotpotQA. The two sets may differ from ChemRxiv Retrieval in text genre rather than subject matter: their corpora are encyclopedic, whereas ChemRxiv Retrieval contains preprint paragraphs. Because retrieval transfer can depend on corpus distribution as well as topic \cite{thakur2021beir}, we looked at whether the two corpora differ in the unmodified base model.

\begin{table}[H]
\centering
\caption{Retrieval performance comparison across chemistry-specific and open‑domain tasks. Dataset descriptions and performance metrics for each model–dataset pair are provided.}
\label{tab:retrieval-wide}
\setlength{\tabcolsep}{5pt}
\resizebox{\textwidth}{!}{
\begin{tabular}{ll ccc | ccc}
\toprule
 & & \multicolumn{3}{c|}{\textbf{Dataset statistics}} & \multicolumn{3}{c}{\textbf{Retrieval Score$^{b}$ $\uparrow$}} \\ 
\cmidrule(lr){3-5}\cmidrule(lr){6-8}
Task & Dataset
& \#Tasks & Avg Queries & Avg Corpus & nomic-embed-text-v1 & \texttt{ChEmbed\textsubscript{vanilla}} & \texttt{ChEmbed\textsubscript{progressive}} \\ 
\midrule
Chemistry Retrieval & ChemTEB(latest)$^{a}$
& 3 & 1682 & $34,153$ & 0.746 & 0.741 & 0.728 \\
Open-domain Retrieval & MTEB Retrieval
& 10 & 1482 & $109,645$ & 0.544 & 0.467 & 0.475 \\
\bottomrule
\end{tabular}}
\par\vspace{2pt}
{\raggedright\footnotesize $^{a}$\,ChemTEB(latest) comprises ChemRxiv Retrieval and two encyclopedic tasks; individual ChemRxiv Retrieval scores appear in Table~\ref{tab:chemrxiv-results}.\par
$^{b}$\,MRR@10 for Chemistry Retrieval; NDCG@10 for Open-domain Retrieval (multiple relevant documents per query). ChemRxiv Retrieval and ChemHotpotQA have one relevant document per query; ChemNQ averages 1.3.\par}
\end{table}

We then asked two questions: whether these two kinds of chemistry text really differ, and what fine-tuning changed. For the first, we encoded 800 ChemRxiv passages and 800 encyclopedic passages with the base model. A logistic-regression classifier told the two groups apart with 95.6\% accuracy, and passage length alone did not account for this. Chemistry literature and chemistry-related encyclopedic text are therefore represented as measurably different populations by the base model. For the second, we measured how close ChemRxiv passages sat to one another before and after fine-tuning. In the base model they were crowded together, with a mean pairwise cosine similarity of 0.570. After fine-tuning this fell to 0.030, so the passages had spread out, a change consistent with the higher retrieval scores \citep{wang2020understanding}. The spreading was stronger for chemistry literature than for encyclopedic text \cite{ethayarajh2019contextual}. Both analyses are described in detail in the Supplementary Material.

Overall, these findings are consistent with distribution-specific adaptation. ChEmbed improves substantially on ChemRxiv, the literature corpus targeted by its training data, while producing lower point scores on the two encyclopedic retrieval tasks. The geometric analyses provide a plausible interpretation of this pattern, but they do not demonstrate that the representation changes caused the differences in retrieval performance. Moreover, ChemNQ and ChemHotpotQA contain only 27 and 18 test queries, respectively. The paired ChemNQ confidence intervals include zero, and the exact paired ChemHotpotQA sign-flip tests give $p=0.125$; consequently, these evaluations do not establish a general loss on encyclopedic retrieval.

We therefore position ChEmbed as a model specialized for chemistry literature retrieval, rather than as a universally superior embedding model for every form of chemistry-related text. More broadly, these results indicate that embedding models should be selected according to the text distribution and retrieval task of the intended corpus, not solely according to a broad domain label such as ``chemistry.''

\subsection{Impact of model fine-tuning on role-based retrieval}
\label{subsec:chebi-role-retrieval}
To demonstrate an example of how this fine-tuning would impact the chemical literature retrieval, we evaluated its impact on a chemical role retrieval task. 
To achieve this goal, a combination of the PubChem dataset \cite{kim2019pubchem} for compound descriptions and ChEBI \citep{malik2025chebi} knowledge graph for structured role annotation was used. 

The hypothesis was that a natural-language query about a chemical role would retrieve PubChem descriptions of compounds that ChEBI directly assigns to that role more effectively. PubChem supplied the descriptions, while ChEBI supplied structured role annotations independently of the embedding models.

\paragraph*{Task Construction}
For each PubChem CID, we retained one 20 to 500 word description and removed exact duplicate texts yielding 44,887 retrieval candidates. We linked these PubChem records to ChEBI using explicit database links or exact full-InChI matches. If several usable PubChem CIDs matched the same ChEBI entity, we retained one CID to avoid duplication. This produced 42,068 ChEBI-to-PubChem mappings involving 40,656 PubChem CIDs. A role-based group contains the distinct mapped CIDs sharing the same direct ChEBI \texttt{has\_role} annotation and does not imply structural relatedness.

Of 1,313 roles with at least one available mapped entity, 597 had 1 to 3 members, 518 had 4 to 30, and 198 had more than 30 members. The primary cohort used the 518 groups with 4 to 30 available entities, containing 5,783 compound-to-role assignments and 4,494 unique entities. We required at least four compounds so that each role could be evaluated as a group. Roles with more than 30 compounds were evaluated separately because they were considerably broader.

For each role, we used three fixed question forms:
\begin{itemize}
    \setlength\itemsep{0em}
    \item ``Which compounds act as [role]?''
    \item ``Find chemical entities with the role of [role].''
    \item ``What chemicals can function as [role]?''
\end{itemize}

With this approach we obtained a total of 1,554 queries. All 44,887 descriptions, were embedded using the baseline Nomic model, ChEmbed\textsubscript{vanilla}, and ChEmbed\textsubscript{progressive}. The resulting embeddings were then ranked against each query. The directly annotated CIDs formed the reference set for that role. Average precision summarizes how early and consistently these reference descriptions appear across the complete ranking, while recall@10 is the proportion of reference descriptions that appear among the first ten results.

\begin{table}[!b]
\centering
\caption{Retrieval of direct ChEBI role members from the 44,887-description PubChem catalog.}
\label{tab:chebi-role-retrieval}
\setlength{\tabcolsep}{4pt}
\begin{tabular}{lcc}
\toprule
Model & Average precision & Recall at 10 \\
\midrule
Nomic & 0.176 & 20.4\% \\
ChEmbed\textsubscript{vanilla} & 0.203 & 22.7\% \\
ChEmbed\textsubscript{progressive} & \textbf{0.315} & \textbf{33.4\%} \\
\bottomrule
\end{tabular}
\end{table}

\begin{figure}[!t]
    \centering
    \includegraphics[width=\linewidth]{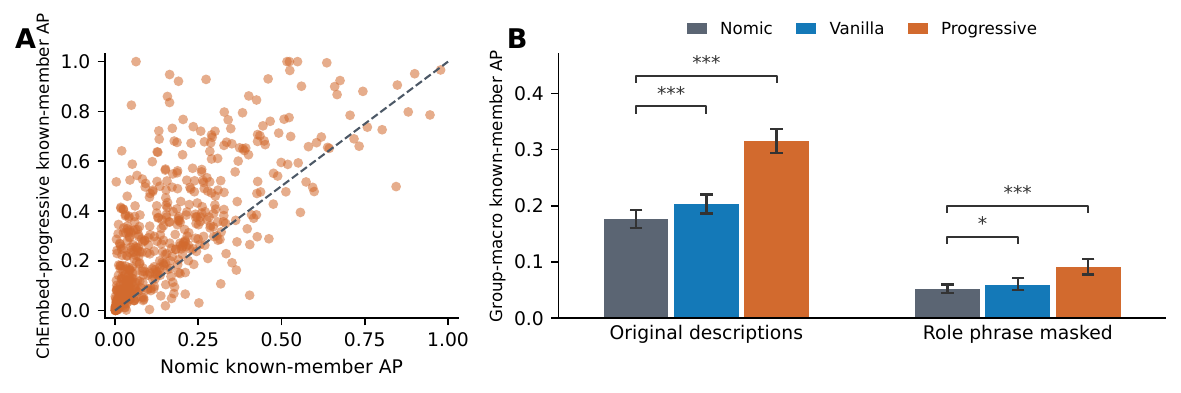}
    \caption{Chemical-role retrieval across the 518-group primary cohort. (a) Group-level average precision (AP) for ChEmbed\textsubscript{progressive} versus Nomic; points above the diagonal had higher average precision in fine-tuned ChEmbed model compared to the base Nomic model. (b) Mean AP over the 518 groups, before and after masking the exact ChEBI role phrase in the descriptions. Error bars are 95\% bootstrap confidence intervals. Stars are paired $t$-tests against Nomic within each condition ($^{*}p<0.05$, $^{***}p<0.001$).}
    \label{fig:role-retrieval-cohort}
\end{figure}

\paragraph*{Findings }
Table~\ref{tab:chebi-role-retrieval} summarizes the results. ChEmbed\textsubscript{progressive} improved performance for 435 of the 518 primary role-based groups relative to Nomic, with a mean gain of $+0.139$ average precision (AP) per group (paired $t$-test across the 518 groups, 95\% CI $[0.124, 0.154]$, $p<0.001$). ChEmbed\textsubscript{vanilla} improved fewer groups, 298 out of 518, and by a much smaller margin. The improvement held across all three query templates, with gains of $+0.168$, $+0.112$ and $+0.137$ AP. A separate sensitivity analysis restricted to one-to-one PubChem-to-ChEBI correspondences produced an essentially identical model advantage, while including groups with more than 30 members preserved the same overall trend. Fig.~\ref{fig:role-retrieval-cohort}-A shows the average precision for each primary role-based group in ChEmbed\textsubscript{progressive} versus the base Nomic model, providing greater transparency than the aggregated AP values reported in Table~\ref{tab:chebi-role-retrieval}. Fig.~\ref{fig:role-retrieval-cohort}-B compares the mean AP of the three models under two conditions: the original PubChem descriptions, and a role-phrase masking control in which the exact queried ChEBI role label was removed from the reference descriptions before ranking. The masking result is discussed below.

To assess how much explicit role wording contributed to the result, we repeated the evaluation after removing the exact queried ChEBI role label from the corresponding reference descriptions. That label appeared in 5,110 of the 5,783 (88\%) evaluated CID-to-role links. After masking it, average precision fell from 0.176 to 0.052 for Nomic and from 0.315 to 0.091 for ChEmbed\textsubscript{progressive}. Explicit role wording is therefore an important part of the retrieval signal. Fig.~\ref{fig:role-retrieval-cohort}-B illustrates these results. A smaller but still significant progressive advantage remained, but this control does not identify its mechanism. Because the role-member descriptions came from a PubChem source used during training, this analysis characterizes learned retrieval behavior rather than held-out chemical knowledge. 
The most strongly supported interpretation is that ChEmbed, particularly the progressive variant, better aligns explicit chemical-role queries with relevant PubChem descriptions. It does not demonstrate inference from molecular structure or learning of a complete chemical ontology.

\begin{figure}[!t]
    \centering
    \includegraphics[width=\linewidth]{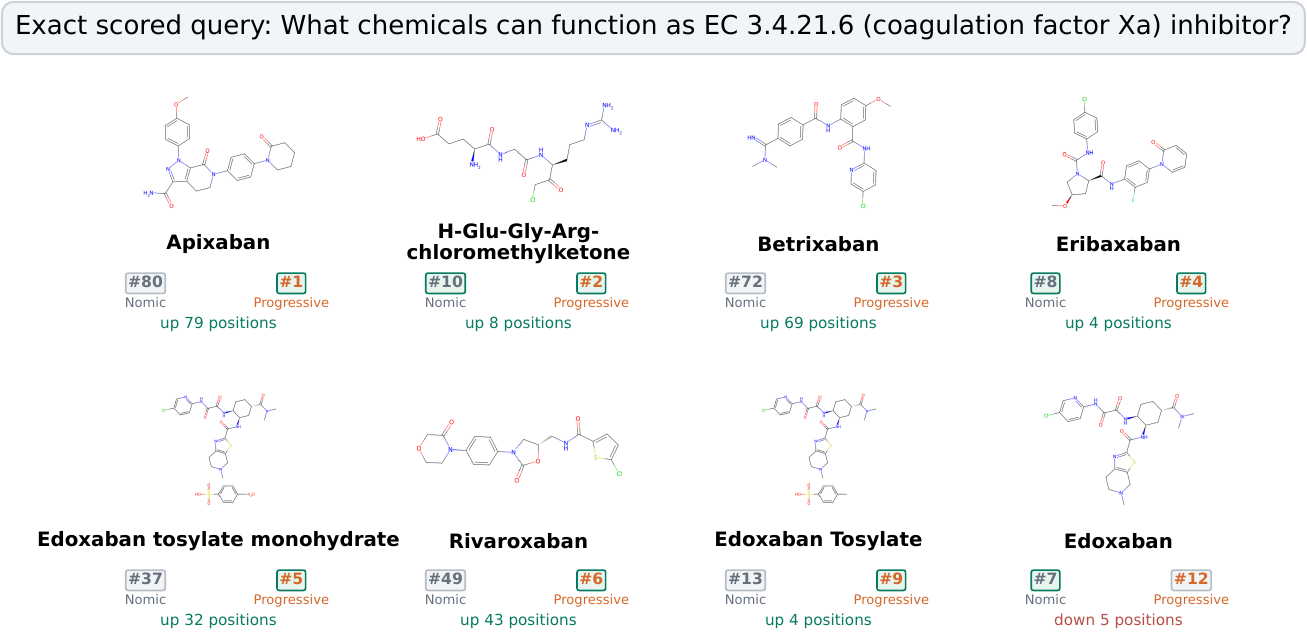}
    \caption{Retrieval ranks for the eight PubChem records directly annotated by ChEBI as coagulation factor Xa inhibitors. For the displayed query, three records appeared among Nomic's first ten results and seven among ChEmbed\textsubscript{progressive}'s first ten. The eight records correspond to six inhibitor identities because edoxaban, edoxaban tosylate, and edoxaban tosylate monohydrate are separate records. Structures identify the records; retrieval used description text only.}
    \label{fig:factor-xa-retrieval}
\end{figure}

To illustrate this result in greater detail, Fig.~\ref{fig:factor-xa-retrieval} shows the eight PubChem records which are defined as coagulation factor Xa inhibitors by ChEBI. The corresponding query in the role-based retrieval evaluation dataset is: ``What chemicals can function as EC 3.4.21.6 (coagulation factor Xa) inhibitor?''. For this query, Nomic embedding retrieved 3 out of 8 directly annotated PubChem entities in its top 10 results; however, ChEmbed\textsubscript{progressive} retrieved 7 out of 8 PubChem entities within the top 10 results. 
It should be noted that the eight PubChem records correspond to six inhibitor identities, since three of these eight entities belong to Edoxaban, which has three records: the free base, the tosylate salt, and the tosylate monohydrate. This example illustrates impact of fine-tuning at the level of individual compounds; however, the overall conclusions presented above are based on the complete set of 518 role-based groups.

\section*{Conclusions}
In this study, we introduce \textbf{ChEmbed}, a family of domain-adapted embedding models specifically designed for retrieving chemical literature. By training on large batches of synthetic query-passage pairs, and by adding chemistry vocabulary under a progressive adaptation schedule, {ChEmbed} substantially improved retrieval accuracy. Our model achieved a meaningful improvement on the target ChemRxiv retrieval benchmark compared to a general embedding model, surpassing larger proprietary models while maintaining efficiency, speed, and supporting long contexts of up to 8192 tokens. Our study revealed important insights into domain adaptation. Specifically, synthetic contrastive training effectively addresses the common issue of data scarcity in specialized fields, such as chemistry. Additionally, while our lightweight vocabulary augmentation strategy does not fully replace complete tokenizer retraining, it proved helpful in practice, most visibly in the role-based retrieval task of Section~\ref{subsec:chebi-role-retrieval}, and offers a pragmatic, efficient alternative when full retraining is not feasible. We also demonstrated the importance of evaluating models on tasks closely matching their intended domain, as illustrated by the differences between ChemRxiv articles and generic benchmarks, such as ChemTEB and MTEB. Nevertheless, some limitations remain. Currently, our model focuses solely on English-language contexts and retrieval tasks, which might not directly generalize to other NLP tasks. Several promising directions can further improve our approach including exploring multilingual capabilities.
Additionally, future efforts could consider building domain-specific embedding models from scratch when sufficient domain data is available, rather than starting from general text embeddings. Regarding tokenization, adopting selective token augmentation guided by domain expertise could overcome the current limitations associated with purely automated WordPiece token selection. Finally, while our methods specifically targeted chemistry, particularly tokenizer augmentation and synthetic query-based contrastive training, they can be broadly generalized to other scientific or technical domains. Beyond generalization, {ChEmbed} directly benefits chemical discovery workflows and literature search tasks by making retrieval more accurate, efficient, and practical for researchers.

\section*{Data Availability}
The following resources are publicly available:
\begin{itemize}
    \item \textbf{Model Weights:} The fine-tuned model weights for all ChEmbed variants are available on Hugging Face at \href{https://huggingface.co/collections/BASF-AI/chembed}{huggingface.co/collections/BASF-AI/chembed} for non-commercial use.
    \item \textbf{Training Data:} A subset of the training data is available at \href{https://huggingface.co/datasets/BASF-AI/ChEmbed-Training}{huggingface.co/datasets/BASF-AI/ChEmbed-Training}. A collection of the raw open-source data used to construct the training corpus is available at \href{https://huggingface.co/collections/BASF-AI/chembed-training-data}{huggingface.co/collections/BASF-AI/chembed-training-data}.
    \item \textbf{Evaluation Benchmark:} The ChemRxiv Retrieval benchmark introduced in this work is available at \href{https://huggingface.co/datasets/BASF-AI/ChemRxivRetrieval}{huggingface.co/datasets/BASF-AI/ChemRxivRetrieval}.
    \item \textbf{Supporting Materials:} The data collection and synthetic query-generation pipelines, evaluation code, raw evaluation results, analyses, expert-review materials, and scripts for the manuscript tables and generated figures are available at \href{https://github.com/HSILA/ChEmbed-Artifacts}{github.com/HSILA/ChEmbed-Artifacts}.
\end{itemize}

\begin{ack}
The authors gratefully acknowledge the financial support received for this research through \textbf{MITACS} funding number IT32409. This project also benefited from the computational resources provided by the Narval clusters of the \textbf{Digital Research Alliance of Canada}. We thank \textbf{Adam Wojciech Bartwiki} for project management, \textbf{Tobias Roth} for support with computational resources and technical assistance, and \textbf{Amir Reza Behzad Moghadam} for serving as the chemistry domain expert in the expert review of generated queries and retrieved passages.
\end{ack}

\bibliographystyle{unsrt}
\bibliography{references}

\clearpage
\section*{Supplementary Material}
\appendix
\setcounter{section}{0}
\renewcommand{\thesection}{S\arabic{section}}
\setcounter{figure}{0}
\renewcommand{\thefigure}{S\arabic{figure}}
\setcounter{equation}{0}
\renewcommand{\theequation}{S\arabic{equation}}
\newcommand{\CEv}{ChEmbed\textsubscript{vanilla}}
\newcommand{\CEp}{ChEmbed\textsubscript{progressive}}
Section 5.3 of the main text presents two related diagnostic analyses. The
first asks whether chemistry literature and chemistry-related encyclopedic
text are already distinguishable in the unmodified base-model space; the
second examines how fine-tuning changes the representations of the same
passages. This supplement gives the design and the evidence for both parts.

\section{Sampling and encoding}
\label{sec:esi-sampling}

We analyzed passages from the three retrieval benchmarks used in the main
paper: ChemRxiv Retrieval, ChemNQ Retrieval and ChemHotpotQA Retrieval. To make
the sampled passages independent at the source-document level, we grouped
ChemRxiv passages by preprint and grouped ChemNQ and ChemHotpotQA passages by
their source encyclopedia page, then kept one passage from each group. The
primary comparison contained 800 ChemRxiv paragraphs and an equally sized
encyclopedic sample composed of 400 ChemNQ and 400 ChemHotpotQA passages;
sampling used seed 1337.

All three models, the \texttt{nomic-embed-text-v1} base encoder and the two
adapted variants, received the same 1{,}600 passages after identical text
preprocessing. Each input began with the model's document-search prefix
\texttt{search\_document:}. The encyclopedic benchmark records store the
source-page title separately from the passage body, so we placed this existing
title field before the passage text, as in the retrieval evaluation. ChemRxiv
records have no separate title field, so we encoded them from the paragraph
text alone. We L2-normalized the resulting 768-dimensional embeddings and ran
inference on CPU in single precision. We use \CEv{} as the primary adapted
model because it isolates the effect of fine-tuning from the tokenizer
augmentation discussed in Section 5.2 of the main text, while \CEp{} provides a
sensitivity check.

\section{Separation before fine-tuning}
\label{sec:esi-separation}

The three corpora all concern chemistry, but they differ in genre, provenance,
formatting and possibly in their topic distributions. We therefore
asked a narrower question: before any ChEmbed fine-tuning, does the
base model represent the literature and encyclopedic samples as measurably
different populations?

We reduced the base-model embeddings to 64 principal components, standardized
them and fitted logistic regression under stratified five-fold
cross-validation. We repeated the dimensionality reduction and scaling inside
each training fold, so information from the held-out passages could not enter
the classifier. The resulting out-of-fold accuracy was 95.6\%
(95\% bootstrap interval 94.6--96.6\%); because the two classes were balanced,
uninformative predictions would be correct only about half of the time. To test
whether the observed accuracy could arise by chance, we repeated the analysis
999 times after randomly shuffling the literature and encyclopedic labels.
None of these shuffled analyses reached the observed accuracy, giving
$p=0.001$. Changing the projection size or cross-validation seed produced
accuracies between 94.9\% and 96.1\%. This use of classification as a two-sample
test follows the general framework of Lopez-Paz and Oquab
\citep{lopezpaz2017revisiting}.

Passage length contributed to the distinction without accounting for most of
it. A probe based only on word and character counts reached 76.1\% accuracy,
reflecting the greater average length of the preprint paragraphs. We therefore
matched literature and encyclopedic passages within ten passage-length bands,
leaving 766 documents with comparable length distributions. On this matched
subset, a classifier using only word and character counts performed at chance
(47.1\%), while the embedding-based classifier remained accurate (93.6\%). A
local analysis led to the same conclusion: among the ten nearest neighbors of
each passage, 86.4\% came from the same corpus class, rather than the
approximately even mixture expected if the classes were interchangeable
\citep{schilling1986multivariate}.

\begin{figure}[H]
\centering
\includegraphics[width=0.8\linewidth]{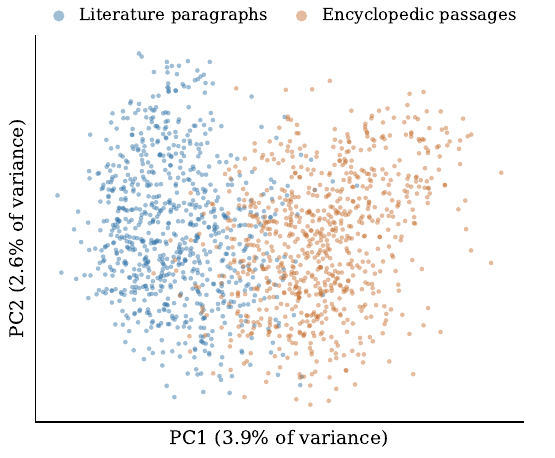}
\caption{Literature and encyclopedic passages in the base model, projected onto
their first two principal components. PCA calculated the projection from the
embeddings alone, and we added the corpus colors afterward. Although the two
components contain only a small share of the total variance, the classes
already occupy substantially different regions.}
\label{fig:esi-separation}
\end{figure}

Figure~\ref{fig:esi-separation} shows this result directly. PCA selected the
two directions that capture the most variation among the passage embeddings
without reference to the corpus labels. Although these two directions contain
only a small part of the total variance, they already reveal substantial
separation.

Together, the classifier, length-matched control and neighborhood analysis show
that the two corpora differ in their base-model representations. They do not
identify the source of that difference, which may combine genre, topic,
formatting and collection provenance.

\section{Changes from fine-tuning}
\label{sec:esi-change}

\begin{figure}[H]
\centering
\includegraphics[width=\linewidth]{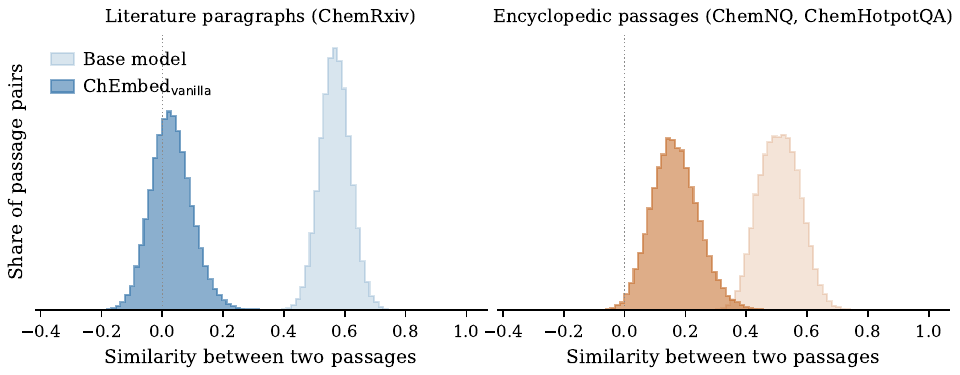}
\caption{Distributions of pairwise cosine similarity before and after
fine-tuning with \CEv{}. Mean similarity falls from 0.570 to 0.030 for the
ChemRxiv literature passages and from 0.517 to 0.163 for the balanced
encyclopedic pool.}
\label{fig:esi-dispersion}
\end{figure}

We next encoded the same passages with the adapted models and compared the
resulting spaces with the base representation. The clearest change concerned
angular dispersion. For ChemRxiv, mean pairwise cosine similarity fell from
0.570 in the base model to 0.030 under \CEv{}, whereas the encyclopedic pool
changed from 0.517 to 0.163. Mean similarity therefore decreased by 0.540 for
ChemRxiv and by 0.354 for the encyclopedic pool: the ChemRxiv decrease was
0.187 greater (95\% bootstrap interval 0.180--0.195). \CEp{} produced the same
qualitative pattern. Fine-tuning therefore dispersed both corpora, more
strongly for the literature passages (Figure~\ref{fig:esi-dispersion}). This
measure describes the concentration of the embedding space; on its own, it does
not show that individual passages became easier to retrieve
\citep{ethayarajh2019contextual}.

The change in local structure was more modest and depended on the scale of the
neighborhood. For each passage, we identified its ten closest passages before
fine-tuning and then counted how many remained among its ten closest passages
after fine-tuning. Under \CEv{}, the average overlap was 35.4\% for ChemRxiv
and 42.7\% for the encyclopedic passages, indicating somewhat greater local
change for ChemRxiv. This difference disappeared when the neighborhood was
expanded from ten to fifty passages (44.0\% against 43.9\%), while the
corresponding \CEp{} comparison reversed direction at the wider scale
(49.7\% against 46.0\%). Linear centered kernel alignment
agreed with the ten-neighbor result \citep{kornblith2019similarity}, but we did
not use it for inference because ordinary bootstrap resampling was strongly
biased at this sample size. These
results support a graded, scale-dependent specialization rather than a
wholesale reorganization confined to chemistry literature.

\section{Limitations}
\label{sec:esi-limits}

These analyses are observational: they establish that the sampled corpora are
distinguishable in the base-model space and that fine-tuning changes their
representations to different degrees, but they neither isolate the corpus
property responsible for the separation nor demonstrate that the geometric
changes caused the retrieval results. The geometry analysis also relies on one
sampling seed, and the two encyclopedic retrieval tasks contain very few test
queries. Their associated score differences should therefore be interpreted as
indicative rather than conclusive.

\end{document}